\documentclass[twocolumn,showpacs,prd,nofootinbib]{revtex4}
\usepackage{amssymb}
\usepackage{enumerate}
\usepackage{hyperref}
\usepackage{dcolumn}

\usepackage{graphicx}
\usepackage{dcolumn}
\usepackage{bm}

\begin{document}


\def\H{{\cal H}}
\def\ttheta{\tilde{\theta}}

\def\beq{\begin{equation}}
\def\eeq{\end{equation}}
\def\bea{\begin{eqnarray}}
\def\eea{\end{eqnarray}}
\def\ben{\begin{enumerate}}
\def\een{\end{enumerate}}
\def\la{\langle}
\def\ra{\rangle}
\def\a{\alpha}
\def\b{\beta}
\def\g{\gamma}\def\G{\Gamma}
\def\d{\delta}
\def\e{\epsilon}
\def\phi{\varphi}
\def\k{\kappa}
\def\l{\lambda}
\def\m{\mu}
\def\n{\nu}
\def\o{\omega}\def\O{\Omega}
\def\p{\pi}
\def\r{\rho}
\def\s{\sigma}
\def\t{\tau}
\def\L{{\cal L}}
\def\S{\Sigma }
\def\gsim{\; \raisebox{-.8ex}{$\stackrel{\textstyle >}{\sim}$}\;}
\def\lsim{\; \raisebox{-.8ex}{$\stackrel{\textstyle <}{\sim}$}\;}
\def\gtrsim{\gsim}
\def\lessim{\lsim}
\def\loc{{\rm local}}
\def\vm{v_{\rm max}}
\def\bh{\bar{h}}
\def\del{\partial}
\def\nab{\nabla}
\def\half{{\textstyle{\frac{1}{2}}}}
\def\fourth{{\textstyle{\frac{1}{4}}}}

\title{A positive energy theorem for Einstein-aether and Ho\v{r}ava gravity}

\author{David Garfinkle}
\affiliation{Department of Physics, Oakland University, Rochester, MI 48309}
\affiliation{Michigan Center for Theoretical Physics, Randall Laboratory of Physics, University of Michigan, Ann Arbor, MI 48109-1120}
\author{Ted Jacobson}
\affiliation{Center for Fundamental Physics, Department of Physics,
University of Maryland, College Park, MD 20742-4111}

\begin{abstract}

Energy positivity is established for a class of solutions to
Einstein-aether theory 
and the IR limit of Ho\v{r}ava gravity
within a certain range 
of coupling parameters. 
The class consists of solutions where the aether 4-vector is 
divergence free on a spacelike surface 
to which it is orthogonal (which implies that the surface is maximal).
In particular, this result holds for spherically symmetric solutions
at a moment of time symmetry. 

\end{abstract}
\maketitle

\section{Introduction}

It is difficult to modify general relativity (GR) in a fashion that meets
basic theoretical requirements such as stability, energy positivity,
and the existence of a well-posed initial value formulation.
Even in the linearized theory one or more of these properties often fails.
Analyzing them in the full, nonlinear theory is of course much
more difficult, so much so that it is rarely done. 

In this paper we
shall establish a fully nonlinear positive energy result for 
two closely related
modifications of GR whose properties and predictions
have been extensively studied over the past several years,
``Einstein-aether theory" and Ho\v{r}ava gravity. 
The first theory, 
called ``ae-theory" for short, 
consists of a dynamical unit timelike vector ``aether" field $u^a$ coupled to
Einstein gravity~\cite{Gasperini:1986ym,Jacobson:2000xp} (for reviews see
\cite{Eling:2004dk,Jacobson:2008aj}).
The vector can be thought of as the
4-velocity of a preferred frame; it
spontaneously breaks local Lorentz symmetry since, being 
a unit vector, it is everywhere non-zero 
in any solution, including flat spacetime.
Ho\v{r}ava gravity~\cite{Horava:2009uw} (for a review see \cite{Sotiriou:2010wn}) 
can be viewed as general relativity coupled to a 
preferred time function $T$, restricted by invariance under 
reparametrizations of $T$. Thus it depends on $T$ only via the 
unit normal (timelike) vector field $NT_{,a}$, with $N=(g^{mn}T_{,m}T_{,n})^{-1/2}$.
Here we consider the so-called ``non-projectable" version of that theory,
in which the lapse function $N$ is allowed to be an arbitrary function of
position, and we include in the action all terms consistent with the symmetry
of foliation preserving diffeomorphisms (so that the theory is dynamically
well behaved~\cite{Blas:2009qj}). 

For both theories, we restrict to terms in the 
Lagrangian with no more than two derivatives of the metric or the vector field.
Ho\v{r}ava gravity theory is then equivalent to a version of ae-theory in which
the aether is restricted to be hypersurface orthogonal at the level of the
action\cite{Blas:2009qj,Jacobson:2010mx}.\footnote{Every hypersurface orthogonal solution 
of ae-theory is a solution of Ho\v{r}ava gravity.  The converse is not true in general, 
but it does hold in spherical symmetry for solutions with a regular 
center~\cite{Blas:2010hb}.} 
Moreover, the total energy of asymptotically flat solutions of the two
theories are given by the same expressions in terms of 
the metric and aether fields (for ae-theory see \cite{Eling:2005zq, Foster:2005fr}, 
for Ho\v{r}ava gravity see \cite{Blas:2011zd, Donnelly:2011df}). 
A positive energy result for 
ae-theory therefore implies a similar result for Ho\v{r}ava gravity. 
We thus focus the discussion on ae-theory.

The Lagrangian of ae-theory depends on four dimensionless coupling
constants $c_{1,2,3,4}$. 
In the hypersurface orthogonal sector of the theory, only the 
combinations $c_{14}=c_1+c_4$, $c_{13}=c_1+c_3$, and $c_2$
enter. The coupling constants of Ho\v{r}ava gravity can be expressed
in terms of these combinations\cite{Blas:2009qj,Jacobson:2010mx}.
Hyperbolicity, stability, and energy
positivity of the linearized theory 
hold for certain ranges of the coupling
constants in ae-theory~\cite{Jacobson:2004ts,Lim:2004js,
Eling:2005zq,Foster:2006az} and Ho\v{r}ava 
gravity~\cite{Blas:2009qj,Blas:2010hb}. 
These ranges coincide in the two theories 
for the spin-2 and spin-0 modes. 
(Ae-theory has an additional spin-1 mode.) 
Here we establish 
a positive energy result
for the full, nonlinear theory.

One might approach this problem by considering the aether field as 
simply one more matter field and trying to use the usual results for 
positivity of mass in general relativity.  
However, the aether Lagrangian
involves the covariant derivative (as opposed to the exterior derivative, 
which is all that is 
needed for minimally coupled scalar fields or for electromagnetism).  
This leads to very different
behavior of the action when the metric is varied and in particular 
to the violation of the dominant energy condition
for the energy-momentum tensor.
Since the dominant energy condition is what is needed for both 
the Schoen-Yau\cite{SY1,SY2}
and Witten\cite{EW,Parker:1981uy} 
proofs of the positive energy theorem, these results do not apply 
to ae-theory, and one might therefore expect that energy is not positive in 
ae-theory. On the other hand, 
a spherically symmetric static vacuum solution is known 
explicitly\cite{Eling:2006df} which has positive
energy despite having everywhere negative aether energy density, 
which suggests that
there may be a general positive energy property.

The question we are addressing here is whether the total energy
$M_{\rm\ae}$ of asymptotically flat solutions in ae-theory is positive. 
(It turns out that this energy is not the same as the ADM mass $M_{\rm ADM}$
that defines the total energy in general relativity, although  
the ratio $M_{\rm\ae}/M_{\rm ADM}$ may be a universal constant.)
As we will show, this is indeed the case for solutions where the vector field is 
hypersurface-orthogonal and where one of those hypersurfaces is asymptotically flat and
has vanishing trace of the extrinsic curvature (i.e., is ``maximal").
In particular, this result holds for spherically symmetric solutions at 
a moment of time symmetry.  The possibility of generalizing this result 
is briefly discussed at the end of the paper.

The method of proof will be to exploit the result of Schoen and Yau~\cite{SY1}
which shows that the ADM mass of an asymptotically flat spatial metric
on an orientable
three-manifold is non-negative if the Ricci scalar is non-negative. 
Although the Ricci scalar of the physical 3-metric is {\it not} generally non-negative,
we will find a conformally related 3-metric with positive Ricci scalar and whose ADM 
mass is equal to $M_{\ae}$ of the original spacetime. 

\section{Einstein-aether theory}
\label{aetheory}

The action for Einstein-aether theory is the most general generally
covariant functional of the spacetime metric $g_{ab}$ and aether
field $u^a$ involving no more than two derivatives (not including
total derivatives),
\beq S = \int \sqrt{-g}~ (L_{\ae}+L_{\rm m})
~d^{4}x \label{action} \eeq
where
\beq L_{\ae} =  \frac{1}{16\pi G} [R-K^{ab}{}_{mn} \nabla_a u^m
\nabla_b u^n +
\lambda(g_{ab}u^a u^b + 1)] \eeq
and $L_{\rm m}$ denotes the matter lagrangian. Here $R$ is
the Ricci scalar, ${K^{ab}}_{mn}$ is defined as
\beq {{K^{ab}}_{mn}} = c_1 g^{ab}g_{mn}+c_2\delta^{a}_{m} \delta^{b}_{n}
+c_3\delta^{a}_{n}\delta^{b}_{m}-c_4u^a u^b g_{mn} \eeq
where the $c_i$ are dimensionless coupling constants, and $\lambda$
is a Lagrange multiplier enforcing the unit timelike constraint on
the aether. The convention used in this paper for metric signature
is $({-}{+}{+}{+})$ and the units are chosen so that the speed of
light defined by the metric $g_{ab}$ is unity.

The field equations from varying (\ref{action}) with respect to
$g^{ab}$, $u^a$, and $\lambda$ are given respectively by
\bea
{G_{ab}} &=& {T^{\ae}_{ab}} + 8\pi GT^{\rm m}_{ab}
\label{EFE}
\\
{\nabla _a} {{J^a}_b} + \lambda {u_b} + {c_4} {a_a} {\nabla _b} {u^a} &=& 0
\label{evolveu}
\\
{u^a}{u_a} &=& -1.\label{unit}
 \eea Here $G_{ab}$ is the Einstein
tensor of the metric $g_{ab}$ and $T^{\rm m}_{ab}$ is the matter stress tensor.
The quantities ${J^a}_b,\; {a_a}$
and the aether stress-energy $T^{\ae}_{ab}$ are given by 
\bea {{J^a}_m} &=&
{{K^{ab}}_{mn}}{\nabla_b}{u^n}
\label{J}\\
{a_a} &=& {u^b}{\nabla _b}{u_a}
\label{a}\\
\nonumber
{T^{\ae}_{ab}} &=& \lambda {u_a}{u_b}+{c_4} {a_a}{a_b}
 - {\textstyle {1 \over 2}} {g_{ab}}{{J^c}_d}{\nabla_c}{u^d}
\\
\nonumber
&+& {c_1} \left ( {\nabla_a}{u_c}{\nabla_b}{u^c} - {\nabla^c}{u_a}
{\nabla_c}{u_b}
\right )
\\
&+& {\nabla _c} \left [ {{J^c}_{(a}}{u_{b)}} + {u^c}{J_{(ab)}} -
{{J_{(a}}^c}{u_{b)}}\right ]. \label{aestress} 
\eea
In the weak-field, slow-motion limit,
ae-theory reduces to
Newtonian gravity with a
value of  Newton's constant $G_{\rm N}$ related to
the parameter $G$ in the
action (\ref{action})  by $G_{\rm N}=G(1-c_{14}/2)^{-1}$~\cite{Carroll:2004ai}.
Note that a sensible Newtonian limit requires that ${c_{14}} < 2$.

The total energy of an asymptotically flat solution, defined in the asymptotic aether
rest frame, is given by
\beq
{M_{\ae}}={M_{\rm ADM}} - {\frac {c_{14}} {8\pi G}} {\int _\infty} {r^a}{a_a},
\label{foster}
\eeq
where ${M_{\rm ADM}}$ is the usual ADM mass (\ref{MADM}),
the integral is over a two-sphere at infinity, and $r^a$ is a unit vector in the radial direction.
(The total energy was first found by Eling\cite{Eling:2005zq} using pseudotensor methods, 
and then by Foster\cite{Foster:2005fr} using Wald's Noether charge method\cite{Wald:1993nt,Iyer:1994ys}. 
It is written in the above form in \cite{Foster:2005fr}.)
At least in the weak-field, slow-motion limit,
we have $G_{\rm N}M_{\ae}=GM_{\rm ADM}$. That is, the difference between
$M_{\ae}$ and $M_{\rm ADM}$ is accounted for by the difference between $G_N$ 
and $G$. We suspect that the equality $G_{\rm N}M_{\ae}=GM_{\rm ADM}$ holds 
in general (i.e.\ not just in the weak field slow motion limit) and therefore that positivity 
of $M_{\rm\ae}$ is equivalent to positivity of $M_{\rm ADM}$ when $c_{14}<2$.  
However, for the purposes of this paper we will only address the question of positivity of 
$M_{\rm\ae}$.

\section{Hypersurface orthogonal case}
We consider here only 
solutions where $u^a$ is hypersurface orthogonal.
This is always the case in spherical symmetry,
but more generally
it is a bona fide restriction.  
On an asymptotically flat slice orthogonal to $u^a$
the spatial metric $h_{ab}$ and extrinsic
curvature $K_{ab}$ are given by
\bea
{h_{ab}} = {g_{ab}} + {u_a}{u_b}
\label{hdef}
\\
{K_{ab}} = - {{h_a}^c}{\nabla _c}{u_b}.
\label{Kdef}
\eea
The trace of the extrinsic curvature is given by
\beq
K=-\nabla_a u^a.
\eeq
If $u^a$ is orthogonal to surfaces of constant $t$ for some function $t$,
then $u_a = N \nabla_a t$ for some ``lapse" function $N$. 
Crucial for our purposes here is the fact that in this case the acceleration
vector $a_a=u^b\nabla_b u_a$ is equal to a spatial 
gradient,
\beq\label{DlnN}
a_a=D_a \ln N,
\eeq
where $D_a$ is the spatial derivative operator.

Using this expression for the acceleration of the aether, the 
total energy (\ref{foster}) becomes
\beq\label{Mae}
{M_{\ae}} = {M_{\rm ADM}} - {\frac {c_{14}} {8\pi G}} {\int _\infty} {r^i} {\partial _i} N
\eeq
where we have chosen
the $t$ coordinate so that $N \to 1$ at infinity. 
Note that this expression for the aether mass is similar to the way that the ADM mass changes
under a conformal transformation.  Consider a conformally transformed metric 
${{\tilde h}_{ab}} = {\Omega ^2}{h_{ab}}$ where $\Omega \to 1$ at infinity.  Since the
ADM mass is given by
\beq\label{MADM}
{M_{\rm ADM}}={\frac 1 {16\pi G}} {\int _\infty} {r^i}({\partial _j}{h_{ji}}-{\partial _i}{h_{jj}})
\eeq
it follows that under a conformal transformation we have
\beq
{{\tilde M}_{\rm ADM}} = {M_{\rm ADM}} - {\frac 1 {4 \pi G}} {\int _\infty} {r^i}{\partial _i}\Omega.
\eeq
Therefore ${M_{\ae}}$ is equal to the ADM mass of a conformally transformed metric,
\beq\label{MaeMtilde}
{M_{\ae}}={{\tilde M}_{\rm ADM}}, \qquad {{\tilde h}_{ab}} = {N^{c_{14}}}{h_{ab}},
\eeq
using the conformal factor $\O=N^{c_{14}/2}$.
The question of whether $M_{\ae}$ is positive thus becomes that of 
whether ${\tilde M}_{\rm ADM}$ is positive. 
 
As in general relativity, the $uu$ component of the Einstein equation
(\ref{EFE}) turns out to be an initial value constraint equation
in the present setting where $u^a$ is orthogonal to the spatial
surface. (This is not a priori obvious, since the 
aether stress tensor (\ref{aestress}) contains second time derivatives. 
The spherical case was treated in detail in \cite{Garfinkle:2007bk},
and a general argument 
is given in \cite{tjprep}.)
This equation reads 
\beq
{^{(3)}}\!R+{K^2}-{K^{ab}}{K_{ab}} = 2 {(T^{\ae}_{ab}+ 8\pi GT^{\rm m}_{ab})}{u^a}{u^b}
\label{hamrel}
\eeq
where ${^{(3)}}R$ is the scalar curvature of the spatial metric.  
Using the fact that $u^a$ is orthogonal to the 
surface, the aether stress tensor (\ref{aestress}) may be 
evaluated as 
\beq
2 {T^{\ae}_{ab}}{u^a}{u^b} =  {c_{14}} (2{D_a}{a^a} 
+ {a_a}{a^a}) - {c_2} {K^2} - {c_{13}}{K_{ab}}{K^{ab}}, 
\label{aerho} 
\eeq 
where $D_a$ is the covariant derivative with respect to the spatial metric.
On substituting equation (\ref{aerho}) into
equation (\ref{hamrel}) one finds  
\bea \nonumber
{^{(3)}}\!R &=&16\pi G\rho+ {c_{14}} (2 {D_a}{a^a}+{a_a}{a^a}) \\
&+& (1-{c_{13}}){K_{ab}}{K^{ab}} - (1+{c_2}){K^2}, \label{hamrel2}
\eea
where $\rho=T^{\rm m}_{ab}{u^a}{u^b}$ is the matter energy density.

\section{Positive energy theorem}
Now if $\rho\ge0$ and $K=0$, then in ordinary general relativity
($c_{1,2,3,4}=0$) this implies ${^{(3)}}R\ge0$, so the theorem
of Schoen and Yau (SY)~\cite{SY1} implies that the ADM energy is positive.
In Einstein-aether theory, provided $c_{14}$ and $1-c_{13}$ are
positive, the $a_aa^a$ and ${K_{ab}}{K^{ab}}$ terms contribute
positively, but the term ${D_a}{a^a}$ has indefinite sign.
Thus, we cannot expect a definite sign for the ADM mass.  However, recall
that it is the aether mass $M_{\ae}$ (\ref{Mae}) 
that is the physical mass of the spacetime, and $M_{\ae}$ is equal to the
ADM mass of a conformally transformed metric (\ref{MaeMtilde}).

Remarkably, precisely the same conformal transformation that 
yields ${M_{\ae}}={{\tilde M}_{\rm ADM}}$ 
removes the indefinite term $D_a a^a$ of (\ref{hamrel2}).
To see this note that the Ricci scalar of $\tilde{h}_{ab}=\Omega^2 h_{ab}$ 
is related to ${^{(3)}}\!R$ by
\beq\label{conformal}
{}^{(3)}\!\tilde{R}=\O^{-2}\Bigl({}^{(3)}\!{R} -4 D^aD_a\ln\O - 2(D^a\ln\O)(D_a\ln\O)\Bigr).
\eeq
With the conformal factor $\O=N^{c_{14}/2}$
we have $D_a\ln\O=(c_{14}/2)a_a$, so
(\ref{hamrel2}) and (\ref{conformal}) together yield
\bea \nonumber
{^{(3)}}\!\tilde{R} &=&N^{-c_{14}}\Bigl(16\pi G\rho+ {c_{14}}(1-c_{14}/2){a_a}{a^a} \\
&+& (1-{c_{13}}){K_{ab}}{K^{ab}} - (1+{c_2}){K^2}\Bigr). \label{hamrel3}
\eea
The result of SY
thus implies that 
the ADM energy of ${\tilde h}_{ab}$ is positive, and therefore the aether mass 
$M_{\ae}$ of $h_{ab}$ is positive, 
provided $\rho\ge0$, $K=0$, $0\le c_{14}\le2$, and $c_{13}\le1$. 

These inequalities on $c_{1,2,3,4}$ are
required by stability and positive energy of the linearized theory. 
What we have found here is that they
also suffice to imply positive energy of hypersurface orthogonal configurations
on maximal slices of the fully nonlinear theory.

The SY theorem holds when the spatial manifold has any
number of asymptotically flat ``ends". This provides a
way to extend the result to the case when the spatial
metric at a moment of time symmetry has a minimal surface.
One can just smoothly join a second copy of the space to itself
along the minimal surface, thus obtaining a space, without the
minimal surface, to which the theorem applies for each end.
Thus the mass of the spherical static vacuum solution, which possesses 
a minimal 2-sphere with a singularity inside, must be positive, 
as indeed it was found to be by explicit construction\cite{Eling:2006df}.
This is an instructive example, since the aether energy density is negative 
everywhere in the solution. 

We now consider how the above result can be generalized.  
One such generalization is to remove the 
condition that $K=0$.  For general relativity this was done by 
Schoen and Yau\cite{SY2} using a 
technique that essentially reduced the problem to one covered 
by their first proof (but required the dominant energy condition, 
which is stronger than the condition $\rho>0$).  
We expect that the method of \cite{SY2} can also be used for 
ae-theory and therefore that the condition that the
slice be maximal can be removed. If so, the theorem could be 
extended to cover in particular time dependent spherical solutions
with $R^3$ topology, 
and possibly spherical black holes\cite{Eling:2006ec,Barausse:2011pu}.

More generally, one might hope to remove the condition that the aether vector field 
is hypersurface orthogonal (this condition always holds 
in Ho\v{r}ava gravity).  
Since the positive mass theorem is essentially a property 
of the constraint equations, to find a general positive mass theorem, one would have 
to examine the general constraint equations in ae-theory.  These equations were 
first written in \cite{Seifert:2007fr}
using a result from \cite{Seifert:2006kv} (see also \cite{tjprep} for a different
derivation).
They are complicated, so
it might be better to start with a simple sub-case, 
such as that of a moment of time symmetry, to see whether a positive energy result 
could be obtained there.  

Finally, it is worth emphasizing that 
a key step in our proof of energy positivity was to express the total
energy for ae-theory in terms of the ADM mass of a particular conformally related spatial metric
whose Ricci scalar is positive under the conditions of the theorem.
It is an unexpected fact that, in the hypersurface orthogonal case, 
the same conformal transformation that makes the energy $M_{\rm\ae}$
equal to the ADM mass ${{\tilde M}_{\rm ADM}}$ removes the indefinite sign
divergence term in the Ricci scalar. 
This may be a hint that a similar conformal transformation 
could be used to generalize the result. 


\vspace{-1mm}
\begin{acknowledgments}
This work was supported in part by the National Science Foundation
under grants  PHY-0601800 and PHY-0903572 through the University of Maryland and
PHY-0456655 and PHY-0855532 through Oakland University.
\end{acknowledgments}

\end{document}